\documentstyle[12pt]{article}
\begin{document}

\baselineskip=13pt plus 0.2pt minus 0.2pt
\lineskip=13pt plus 0.2pt minus 0.2pt

\begin{flushright}
hep-ph/9509336 \\
LA-UR-95-2776 \\
\end{flushright}

\begin{center}
\Large{\bf Antimatter Gravity\\
and\\
Antihydrogen Production\footnote{Email: pbar@cernvm.cern.ch;
goldman@hotelcal.lanl.gov; mmn@pion.lanl.gov}}

\vspace{0.25in}

\large

\bigskip

Michael H. Holzscheiter,$^{\dagger}$
T.  Goldman,$^*$
and\\
 Michael Martin Nieto$^{*+}$\\
\end{center}

\vspace{0.2in}

\begin{flushleft}
{\it
$^{\dagger}$Physics Division, Los Alamos National Laboratory,
University of\\
$~~~$California, Los Alamos, New Mexico 87545, U.S.A.\\
$^*$Theoretical Division, Los Alamos National Laboratory,
University of\\
$~~~$California, Los Alamos, New Mexico 87545, U.S.A.\\
$^+$Abteilung f\"ur Quantenphysik, Universit\"at Ulm, D-89069 Ulm,\\
$~~~$GERMANY}
\end{flushleft}
\normalsize

\vspace{0.2in}
\begin{center}

{ABSTRACT}

\end{center}

\begin{quotation}

Certain modern theories of gravity predict that antimatter will fall
differently than matter in the Earth's gravitational field.  However, no
experimental tests of gravity on antimatter exist and all conclusions drawn
from experiments on matter depend, at some level, on a specific model.
We have proposed  a direct measurement that would compare the gravitational
acceleration of antiprotons to that of negatively charged hydrogen ions.
Substantial progress towards the development of this experiment has been
achieved.  Based on our work a number of alternative proposals for measuring
``$g$" on both charged and neutral antimatter have been made. We summarize
the present status of our experiment and also discuss the steps that would be
necessary to produce antihydrogen in an environment suitable for gravity
measurements.

\vspace{0.25in}

\end{quotation}

\vspace{0.3in}

\newpage

\section{Introduction}

An area of experimental physics which needs attention is the study
of the gravitational acceleration of antimatter compared to that of matter. A
number of attempts to unify gravity with the electro-weak and strong
interactions call into question
 the general belief that the results
should be identical.  However,
 no experimental tests have been performed \cite{GRAVITY}.

We have proposed an experiment to study gravity on antiprotons
\cite{PS200} and this
experiment is in progress at LEAR/CERN in Geneva, Switzerland. Two main
technical problems can be identified in this proposal.
Firstly, one needs a sufficiently large number of ultra-cold
antiprotons to obtain a statistically significant result.  Secondly,
electromagnetic forces acting on the charged particles need to be reduced
to an extremely low level. To make the force of gravity observable
one must
reduce electric fields to a level well below $10^{-7}$ V/m in the critical
portion of
the experimental set-up. (One must also minimize
any magnetic gradient fields).
Such a reduction seems possible, but very difficult.
This difficulty has prompted
alternative proposals to utilize a neutral particle, specifically the
antihydrogen atom. Both these approaches are being
pursued and are covered by this work.

The ballistic ``Galileo" method to measure gravity
on antiprotons is described in detail in reference \cite{PS200a}.
It uses the idea originated by Witteborn and
Fairbank  \cite{wf} to measure the gravitational acceleration
of electrons. One launches a large number of ultra-slow particles
upwards against the gravitational force and  measures the time-of-flight
for each individual particle. The resulting time-of-flight distribution
will exhibit a cut-off time representing the fact that particles below a
certain initial kinetic energy cannot climb up to the height of the detector.
A simple calculation shows that this cut-off time ``$\tau$" is directly
dependent on
the gravitational acceleration ``$g$"and on the geometrical length ``$L$"
of the experiment, but does not depend on the
intrinsic kinematic parameters of the particle sample:
\begin{equation}
\tau = \sqrt{L/2g}~.
\end{equation}

The plan using a neutral particle would proceed
via the production of a composite particle, the antihydrogen atom,
consisting of
an antiproton and a positron. To measure gravity (as well as to provide the
opportunity of precision spectroscopy for CPT tests at unprecedented levels of
accuracy) would require these antihydrogen atoms to have extremely low kinetic
energies and they would
preferably also be confined in a small volume in space (in a trap
configuration). Two proposals to achieve this have been discussed.

The first
approach uses a
recombination process between antiprotons in a Penning trap and a dense
positron plasma. (We will discuss it in Sec. 4.1).
The second method, under consideration
 by our collaboration, is a charge-exchange reaction
between antiprotons and positronium atoms.
Even with a reduced production rate there are
advantages to this approach. The cross section exhibits a broad
maximum
over an energy range from 0 to 20 keV, so in principle the antiprotons do not
have to be ultra cold.
More importantly, antihydrogen is produced directly in the
ground state and low-lying excited states. Therefore, the system is not as
susceptible to field ionization
as the high-{\it n} states produced with the first method
and is furthermore directly accessible to
spectroscopic measurements. Specific states can be preferentially populated
using excited positronium states, which also increases the reaction rate by the
fourth power of the principal quantum number, $n$, of the positronium atoms.

The following lays out our basic experimental approach and describes the
progress made to date.


\section{Capture and cooling of antiprotons}

An important prerequisite for both the gravity measurement on antiprotons and
also the production of antihydrogen is a sufficient supply of cold antiprotons.
Antiprotons are produced in high-energy collisions, and are typically injected
into large accumulator rings (at CERN and Fermilab) to be used in high-energy
physics experiment. At CERN, a portion of the antiprotons is transferred to
a low-energy antiproton ring (LEAR), where they are made available
to physics users at energies
as low as 5.9 MeV. To achieve the
extremely low energies necessary for the
work discussed here, approximately 10 orders of magnitude have to be bridged
with an
efficiency of a few tenths of a percent or better.

We have developed a method
based on a combination of energy loss in material and electrodynamical capture
of charged particles in a Penning trap
(which is a superposition of a strong homogeneous
magnetic field and an electrical quadrupole field).
5.9 MeV particles are passed
through a thin target foil and, with a proper choice of target thickness, up to
5 percent of the antiprotons emerge from the down-stream side of the target at
energies below 50 keV. These ``low-energy antiprotons" are dynamically captured
in our large ``catching" Penning trap.  This is done
by rapidly switching the electrical potential on the
electrodes once the beam pulse from the accelerator has fully emerged from
the target. In our set-up at LEAR we have demonstrated the direct capture of up
to one million antiprotons in a trap specifically designed to match the LEAR
output phase space. Once captured, these antiprotons are cooled by
interaction with cold electrons (which have been preloaded into the trap and
have cooled by synchrotron radiation before the beam pulse arrives).

Recently we
have demonstrated the successful collection of 60 percent of the initially
captured antiprotons into a small, harmonic region of our large trap at
energies below 1 eV.
Surprisingly, the observed storage time of the cold antiprotons in this
trap was in excess of one hour, even though the residual gas pressure was as
high as $10^{-11}$ Torr.
In these runs the annihilation of the antiprotons on the residual gas molecules
was monitored with external scintillators. After an initial cool down period,
no annihilation signals were seen above the background cosmic-ray events, even
though a significant number of antiprotons was still present in the trap at
the time of release one hour after capture \cite{prl}.
This  enables us to now consider the next stages of
experimentation, both with charged antiprotons and neutral antihydrogen.


\section{The Antiproton Gravity Experiment}

The principle of the proposed method to measure the gravitational
acceleration of antiprotons was pioneered in a challenging experiment
performed at Stanford University in the 60's \cite{wf}. The authors of this
experiment
reported a zero net force on free electrons traversing a vertical shield tube.
This result was explained by the fact that the free electron gas in the metal
tube (used to shield external electric fields) sagged under the influence of
gravity.  Thereby an electric field was produced
 which counteracted the force of gravity. To do this measurement,
stray electric fields from variations in the work function along the inside
surface of the shield tube (the Patch effect) had to be below $10^{-11}$ V/m.

In the experiment proposed here the requirements on reducing the field
inside the tube can be relaxed by three orders of magnitude due to the inertial
mass of the antiproton being 2000 times larger than the electron mass.
Even so, to minimize the electric
fields on the central axis of a cylinder due to patches with differing work
functions along the surface of the inside wall of the cylinder,
we have conducted a series of systematic studies to
minimize the variation in work function from patch to patch as well as the size
of the individual patches.

Patch sizes can either be dominated by the intrinsic
structure of the material or by
substances that adsorb on the surface. For these reasons it
became apparent that an amorphous, non-reactive, single component (no alloys)
material is the best choice. We have used a vibrating capacitive probe (Kelvin
probe) to study small samples of different surfaces and have identified two
candidates: graphite in the form of an aerosol spray (Aerodag) and ultra-thin
layers of gold on a germanium sub-layer \cite{jordan1}. Both these surfaces do
not exhibit
any work function variations at the level of the instrument resolution of the
Kelvin probe, 1 mV.  But the observed changes in overall work function when
using graphite samples with different degrees of orientational disorder may
point towards work function variations not too far below the resolution
limit \cite{jordan2}. Such an effect is not expected for the gold/germanium
surfaces. Work in this area is now continuing at our collaborating institution
in Genova, Italy, with an improved instrumental set-up. Here a probe with a
much larger area will be used, allowing us to study large-scale variations of
the workfunction \cite{Vittorio}.

One obvious method to dramatically increase the sensitivity of
these studies is to
use the time-of-flight method itself as a probe for the surface electric fields
present in the system. For this purpose we have constructed a test experiment
consisting of an ion source, a drift tube, and a particle detector \cite{mo10}.
Ions are
transferred from the ion source to the entrance of the drift tube at 1 -
2 keV energy to minimize the error in the definition of the start time. Here
they are retarded by an electric potential to zero mean kinetic energy. Some
ions are rejected and
others still traverse the drift tube at high energies.
With the
proper choice of the current density in the pulse from the ion source,
a small portion of ions (preferably less than one per pulse) enters the
drift tube near zero kinetic energy. These particles are sensitive to
the electric field along the tube axis and their time-of-flight
will provide a measure of the rms fluctuations of these fields.

In preliminary tests of this system we established an energy spread of the ion
source output of less than 7 eV. This, in conjunction with the output current,
would yield a sufficient density of particles in the low-energy bins of
interest. But these tests also revealed a high neutral
background density from the ion source which is saturating the particle
detector during and shortly after the ion pulse is injected, therefore blanking
the signal of late arrivals. This prompted us
to install a small Penning trap as an intermediate storage unit between the ion
source and the drift tube. This trap can dynamically capture more than $10^6$
charged particles from the ion source and release them again with a delay
sufficiently long to let the neutral background decay away. The installation of
this trap is currently under way and first results are expected soon.


\section{Antihydrogen formation}

\subsection{Background}

A variety of schemes to produce
antihydrogen  have been proposed in the literature \cite{mo13}. Amongst these,
only the reactions listed below (with the appropriate references
for detailed discussions) will yield the ultra-low energy antihydrogen atoms
needed:
\begin{equation}
e^+  + e^+ + \overline{p} \Rightarrow \overline{H} + e^+ + h\nu   \label{gH}
\end{equation}
(see Ref. \cite{mo14}), and the two  reactions
\begin{equation}
Ps + \overline{p}  \Rightarrow  \overline{H} + e^- + h\nu~,
\label{h1H}
\end{equation}
\begin{equation}
Ps^* + \overline{p}  \Rightarrow  \overline{H} + e^- + h\nu ~.
\label{h2H}
\end{equation}
(See Referenences
\cite{mo15}-\cite{mo17} and
\cite{mo18}-\cite{mo19}, respectively.)

In the first case, both constituents forming antihydrogen need to be trapped.
In the last two cases, only antiprotons need to be confined before the
recombination process since the positron is delivered in form of a positronium
beam. Both methods have distinct advantages and disadvantages and, depending on
the final application, either one could be a better choice.

In the first process,
the positron density provides a second positron in the
vicinity of a collision between antiprotons and positrons to assist in the
conservation of energy and momentum.
This makes
the rate constant  strongly temperature dependent:
\begin{equation}
\Lambda = 6 \times 10^{ -12} ( 4.2/T )^{9/2}  n^2_e  s^{-1} ~.
\end{equation}
Therefore, the rate
benefits vastly from cooling the particles. Because of the mass
difference between positrons and antiprotons, with positrons at 4.2 K the
antiprotons could have energies as high as 1 eV before the recombination rate
is significantly affected. This could be used to form a beam of
antihydrogen atoms,
at eV energies, leaving the trap in the
axial direction.  (However, the beam
quality would strongly depend on how well the axial energy of the antiprotons
could be decoupled from the radial energy.)

While the theoretical production rate
for this process appears, at first sight, to
 be extremely high (with a $10^7$/cm$^3$ positrons at 4.2 K
one obtains $\Lambda = 6 \times 10^6$ s$^{-1}$),
two critical problems have been identified:

Firstly, the antihydrogen atoms are created
in an extremely high Rydberg state ($n = 100$ or larger). This gives
rise to the possibility that
these loosely bound systems are field ionized by the electric field
gradients present in the trap, needed to store the antiprotons and positrons
prior to
recombination.
This may partially be the reason that  the
charge conjugate reaction (protons on electrons producing hydrogen atoms) has
not yet been observed in recent attempts \cite{Gabrbled}.
Secondly, the neutral atom traps
available for antihydrogen only stabilize specific spin states (low-field
seeking states) and a spin change during de-excitation form these high $n$
levels to the ground state must be carefully avoided.

To combat the ionization problem,
a mixture of collisional and spontaneous
deexcitation might be used,
provided the antihydrogen does not drift outside the positron plasma on the
same time scale. Alternatively, laser induced de-excitation could be attempted.
Additionally, magnetic field effects both on the recombination process and on
the survival of the antihydrogen Rydberg atoms need to be studied in detail
before a final assessment of the advantages and the disadvantages
of this method for a
specific application can be made.

Alternatively to the process of Eq.(\ref{gH}),
one can enhance the radiative antihydrogen formation rate by
several orders of magnitude by coupling the recombination process to a third
particle (for energy and momentum conservation) using collisions between
positronium atoms and antiprotons \cite{mo19}.
[See Eqs. (\ref{h1H}) and (\ref{h2H}).] This process can
be interpreted as Auger
capture of the positron to the antiproton.  Cross sections have been
estimated by Humberston, et al. \cite{mo16},
using charge conjugation and
time reversal
to link the cross section for positronium formation in collisions between
positrons and hydrogen to the antihydrogen formation cross sections. Early
calculations assumed both antihydrogen and positronium to be in the ground
state, resulting in a production cross section of  $3.2 \times 10^{-16}$ cm$^2$
with a broad
maximum at a $\overline{p}$ energy of 2.5 keV. Calculations of the
total antihydrogen
formation cross section using classical and semi-classical methods \cite{mo21}
have
obtained values  which are considerably larger than the ground-state results.
Values for the formation of  antihydrogen in excited states are given by
Ermolaev, et al. \cite{mo22} and indicate that there is a large cross section
to
low-lying excited states.

\subsection{Recombination experiments with protons}

To test the validity of the calculated cross-sections,
our collaboration has set up an experiment  to perform the
charge-conjugate experiment of forming hydrogen atoms via collisions between
protons and positronium \cite{charlton}.
Besides testing the theoretical predictions for production, this experiment
also would allow
 us to develop the necessary technology of positronium production
and handling.

In this experiment a pulsed, low-energy positron beam is made to impinge upon a
heated silver target.  This acts as a high-vacuum source of positronium (Ps)
atoms. An intense proton beam
($100-200~\mu$A at $9$ keV) crosses the Ps target.  The
production of a hydrogen atom is signaled by the detection of the
(low-energy) fragment $e^+$ in time relation to the initial $e^+$ pulse.

The initial low-energy $e^+$'s are produced using a rare-gas solid moderator
bombarded by $\beta^+$ particles from a $3$ GBq ($80$ mCi) $^{22}$Na
 radioactive
source.
The slow $e^+$'s are accelerated to $300$ eV in the axial confining field
of around $10^{-2}$ Tesla.
Using a pair of curved
{\bf E$\times$B} plates, the $e^+$'s are also deflected by $25$ mm
 to remove the remainder of the
beam line from the direct
line-of-sight of the radioactive source. The $e^+$'s are then decelerated to
$10$ eV and passed
into a buncher. At this point the d. c. beam has an intensity
of around $5 \times 10^6~e^+/$s. The potential in the one-meter-long buncher
varies quadratically and, if switched on, produces a time-focussed ejected
$e^+$ pulse. Currently the bunching efficiency is approximately $10$ \%,
 in good
agreement with  expectations based upon the $120$ ns pulse width and the
$100$ kHz repetition rate. The width of the positron pulse has been measured
to be below $5$ ns. Provisions to accumulate positrons in the buncher for a
higher-intensity output have been made but not yet tested. The positrons
would
leave the buncher at a kinetic energy of $7.5$ keV and impinge upon a $200$ nm
Ag$(100)$
foil.  There  they would be converted with a $10 - 20$ \% efficiency to
low-energy positronium atoms.

The proton beam is brought to a focus about $2$ mm in front of the
Ag$(100)$
foil to produce the optimum overlap between the positronium and the protons.
The fragment $e^+$ resulting from the production of a hydrogen atom is
accelerated by applying   $600$ V across the $20$ mm gap between the Ag foil
and the first element of the extraction optics. It is then deflected through
two $90^o$ bends, which are finely tuned to allow passage of the $e^+$ to the
channel electron-multiplier array (CEMA) detector. This detector is placed in
coincidence with a NaI(Tl) detector to unambiguously
identify the $e^+$ signal.

The individual parts of this experiment have been constructed and tested.
Our efforts are
now going towards testing the interfacing of these individual
components. By using the calculated cross section ($6 \times 10^{-16}$ cm$^2$),
beam strengths ($5 \times 10^6~e^+/$s and $150~{\mu}$A proton current), the
production efficiency of Ps in transmission from the heated Ag foil ($10$ \%),
the ortho-Ps
lifetime ($142$ ns), and realistic detection and bunching
efficiencies, the estimated event rate will be around $10^{-2}/$s.  This
should be clearly distinguishable above the background.

After the conclusion of this experiment, the technologies of producing
positrons, injecting them into a target chamber, and converting them to
low-energy positronium
atoms can be implemented into our trapping experiment at
CERN. A detector arrangement consisting of an array of small silicon pads is
being designed.
This detector array will have enough spatial resolution to
be able to discriminate between the pions from
antiprotons annihilating on the residual gas in the trap and those resulting
from neutral antihydrogen atoms escaping from the electromagnetic confinement
of the Penning trap and annihilating on a nearby target.

The first goals of the antihydrogen
experiment will  be to detect the production of antihydrogen atoms using
the positronium-antiproton reaction and to verify that these antihydrogen
atoms are stable against the electric field gradients present in the Penning
trap environment. In a second stage of the experiment we propose to detect the
Lyman-$\alpha$ radiation from antihydrogen atoms
formed directly  in the n=2
state, thus
verifying the population of low-lying states. Once this has been
completed, physics experiments with antihydrogen atoms can be considered.

\subsection{Possible experiments with antihydrogen}

Considering the effort necessary to produce antihydrogen one must naturally ask
the question what the physics benefits of such an endeavor would be. In
principle, these can be found in two areas: 1) Comparison of results of
spectroscopic measurements of hydrogen and antihydrogen, which would constitute
a test of CPT at a level rivaling even the result on the kaon system, and 2)
the
study of the gravitational interaction of antimatter with the Earth's
gravitational field, which
would test the validity of the weak equivalence principle
(WEP) and possibly shed light on the problem of unifying gravity with the
three other forces.

Over the last decade, the
 precision of spectroscopic studies of hydrogen advanced enormously.
 Today the highest precisions have been achieved for the hyperfine
structure ($6.4\times10^{-13}$) and for the 1$s$-2$s$ transition
($1.8\times10^{-11}$). Based on the
lifetime of the 2s state (1/8 second) and the natural linewidth connected to
this, a possible precision of $10^{-18}$ has been theorized.
This latter precision would
require using trapped hydrogen atoms, an environment which would be
directly applicable to antihydrogen.

Currently the best tests of CPT invariance
have been performed in the kaon system followed by precision comparisons of
the magnetic moments and masses of the electron, positron, proton,
and antiproton. The
comparison of the inertial masses of the
proton and the antiproton have now reached a
precision of $1.1\times 10^{-9}$ \cite{gabrielse}.
But in the strict sense, this must be considered only
a measurement of the ratio of the charge-to-mass ratios of the two particles.
This  needs to be combined with the measurement of the Rydberg of
protonium to extract a
independent CPT test \cite{hughes}.
With  the current precision on this quantity \cite{robertson}, a
CPT test of only $2\times10^{-5}$ is possible.
Using the Rydberg of antihydrogen, one can
construct a limit for the charge equality between antiproton and proton which
is entirely based on frequency measurements, and could yield a direct
test of CPT at the level of $10^{-11}$.

Simple ballistic measurements of $``g"$ on antihydrogen are difficult and
unlikely to yield a very high precision measurement because of
the photon recoil limit of approximately $2.4$ mK. Consequently, a more
elaborate
method, using a horizontal beam of ultra-low energy antihydrogen
atoms, has recently been suggested \cite{duke}.
Here the vertical deflection of the beam
would be measured using a transmission interferometer. The approach is
based on the example of an interferometric measurement of a beam of cold
sodium atoms \cite{keith} in which the phase of the interference pattern was
obtained
to $0.1$ radian with only 4000 atoms in the beam.

Assuming an antihydrogen beam
with a velocity of $v = 10^4$ m/s (which corresponds to a wavelength of 40 pm)
and a deflection of $0.8 \mu$ m, an uncertainty in the phase measurement of
$0.1$ radian would lead to an uncertainty in the measurement of $``g"$ of
$1\%$.
One needs to realize that such an experiment would not require trapping of the
antihydrogen atoms and could therefore be considered a first stage experiment
comparable to, but not vastly superior to,  the ballistic measurement of
experiment PS200.

If the formed antihydrogen atoms could be trapped and laser cooled to form
an atomic fountain, a potentially much more powerful
method could be developed based on the work of
by  Chu and collaborators \cite{chu}.
In their experiment they  used velocity-sensitive,
stimulated Raman transitions to measure the gravitational
acceleration, $``g"$, of laser-cooled sodium atoms
in an atomic fountain geometry.
An ultra-cold beam from an atomic trap was launched upwards and
was subjected to
three subsequent pulses to drive a two-photon Raman transition between the
F = 1 and 2 levels in the $^{3}S_{1/2}$ state. A first ($\pi/2$)
pulse  prepared the
sample in a superposition of the two states, the second
($\pi$) pulse  reversed the populations, and the third  ($\pi/2$)
pulse brought
the wave packets to interference. The interference was  detected by probing
the number of atoms in state 2.

In the absence of external forces acting
on the atoms the final state of an atom will depend on the phase of the driving
Raman field. In the frame of reference falling with the atom, the Raman
light fields appear
linearly Doppler-shifted  in time, which shows up as a phase
shift varying as the square of the time.
 Using a $50$ ms delay between the pulses, distinct interference fringes were
observed, and a least square fit to the data gave an uncertainty in the phase
determination of $3\times 10^{-3}$ cycles.
This represented a sensitivity to $``g"$  of $\delta g/g = 3 \times 10^{-8}$.

Despite the enormous advances in the field of
hydrogen
spectroscopy over the last years, hydrogen (and certainly antihydrogen)
is ill-suited
for high
precision measurements. A translation of the above method to the
case of antihydrogen will not be trivial and straight forward. A large
problem will be imposed by the much higher photon recoil limit for laser
cooling hydrogen atoms ($\approx 3$ mK) which gives an rms velocity spread of
approximately 700 cm/sec. A much faster fountain beam, resulting in greatly
increased experimental dimensions, will have to be used. Therefore, a much
larger fraction of the initial beam pulse will be lost due to ballistic
spreading during the flight time of the sample. Much less than $1~\%$ of the
initial population can be expected to contribute to the fringes.
Nevertheless, this method
is the only one identifiable in the current literature which shows the
potential of a high precision measurement of $``g"$ on antihydrogen atoms.

\section{Summary}

Recent progress in trapping and storing low-energy antiprotons has created
exciting opportunities for fundamental research, especially in the areas of
CPT violation and gravity. We propose to explore what, in our opinion, is the
most promising route to produce antihydrogen in both an intrinsic state and an
external environment suitable for high-precision measurements in these areas.
 Success in
this activity could help in convincing the CERN leadership to extent
the lifetime of the LEAR program at CERN beyond the current shut-down date at
the end of
1996
as well as in convincing other laboratories worldwide (FNAL, BNL, KEK) to add
ultra-cold antiprotons to their menu.

\end{document}